\documentclass{article}
\usepackage[]{psfig}
\newcommand{\be}{\begin{equation}}
\newcommand{\ee}{\end{equation}}
\newcommand{\ba}{\begin{eqnarray}}
\newcommand{\ea}{\end{eqnarray}}
\newcommand{\hp}{\,\hat{\!\bar \phi}}
\begin{document}
\title{\bf Self-dual Chern-Simons solitons in noncommutative space}
\author{G.S.~Lozano\thanks{Associated with CONICET}\,,\\
{\normalsize\it Departamento de F\'\i sica, FCEyN, Universidad de
Buenos Aires}\\ {\normalsize\it Pab.1, Ciudad Universitaria,
Buenos Aires,Argentina}\\ ~
\\
E.F.~Moreno$^*$ \, and \,
F.A.~Schaposnik\thanks{Associated with CICBA}
\\
{\normalsize\it Departamento de F\'\i sica, Universidad Nacional
de La Plata}\\ {\normalsize\it C.C. 67, 1900 La Plata, Argentina}}

\maketitle

\begin{abstract}
We construct exact soliton solutions to the Chern-Simons-Higgs
system in noncommutative space, for non-relativistic and
relativistic models. In both cases we find regular vortex-like
solutions to the BPS equations which approach the ordinary
selfdual non-topological and topological solitons when he
noncommutative parameter $\theta$ goes to zero.
\end{abstract}
\date{}

%


\newpage

\section{Introduction}
After the connection between string theory and noncommutative
field theories was unraveled \cite{CDS}-\cite{SW}, the study of
solitons and instantons in noncommutative spacetimes has attracted
much attention \cite{HA}-\cite{LMS}. Chern-Simons (CS) theories in
commutative space have played a central role for the understanding
of relevant phenomena in planar physics \cite{Fr},\cite{J} and
some of their properties started to be explored recently in the
noncommutative case \cite{Chu}-\cite{JMW2}.

In ordinary $2+1$ dimensional space, models of relativistic and
non-relativistic matter minimally coupled to gauge fields whose
dynamics is governed by a CS term have self-dual vortex-like
solutions\cite{JP}-\cite{JW}. It is then interesting to determine
if this kind of solutions are also present in the non-commutative
extension of these models. In this work we shall study the
existence and properties of vortex-like solitons for Chern-Simons
matter systems in noncommutative $2+1$ dimensional space. Our
approach follows closely that developed in \cite{LMS} for
constructing exact noncommutative vortex solutions in the
Maxwell-Higgs system except that now the dynamics of the gauge
field is governed by a CS Lagrangian. Also, we consider the
nonrelativistic case introduced in \cite{JP} in view of its
relevance for studying Bohm-Aharonov effect and other interesting
phenomena in planar physics.

After introducing the noncommutative models in section 2, we
derive BPS equations and construct  explicit solutions in the
nonrelativistic case in section 3. We find  a family of
non-topological BPS solitons parametrized by a constant $f_0$. The
corresponding magnetic flux $\Phi$ is in general non quantized but
becomes an integer in the $\theta \to 0$ limit. In section 4 we
consider  the relativistic case, and construct BPS topological
solitons with quantized magnetic flux which coincide with the
regular ordinary vortex solutions when $\theta \to 0$. Finally, we
present our conclusions in section 5.

\section{The model}

We consider 3-dimensional space-time with coordinates $X^\mu$ ($i
= 0,1,2$) obeying the following noncommutative relations
\be
[X^\mu,X^\nu] = i \theta^{\mu\nu} \label{00}
\ee
The real antisymmetric matrix  $\theta^{\mu\nu}$, can be brought
into its canonical (Darboux) form by an appropriate orthogonal
rotation

\be
[X^1,X^2] = i\theta \, , \;\;\;\;\; [X^1,X^0] = [X^2,X^0] = 0
\label{1} \ee

One way to describe field theories in noncommutative space is by
introducing a Moyal product $*$ between ordinary functions. To
this end, one can establish a one to one correspondence between
operators $\hat f$ and ordinary functions $f$ through a Weyl
ordering
\be
\hat f(X^1,X^2) = \frac{1}{2\pi} \int d^2k \tilde f(k_1,k_2)
\exp\left( i(k_1X^1 + k_2X^2) \right)
\ee
Then, the product of two Weyl ordered operators $\hat f \hat g$
corresponds to a function $f*g(x)$ defined as

\be
f*g(x) =
\left.\exp\left(\frac{i\theta}{2}(\partial_{x_1}\partial_{y_2} -
\partial_{x_2}\partial_{y_1}) \right) f(x_1,x_2) g(y_1,y_2) \right
\vert_{x_1=x_2, y_1 = y_2}
\ee

Given a $U(1)$ gauge field $A_\mu(x)$, the field strength
$F_{\mu\nu}$ is defined as
\be
F_{\mu\nu} = \partial_\mu A_\nu - \partial_\nu  A_\mu - i(A_\mu *
A_\nu - A_\nu * A_\mu) \ee
We shall couple the gauge field to a complex scalar field $\phi$
with covariant derivative
\be
D_\mu \phi = \partial_\mu \phi -i A_\mu * \phi \ee
An alternative approach to noncommutative field theories which has
shown to be very useful in finding soliton solutions \cite{GMS}
is to directly work with operators in the phase space $(X^1,X^2)$,
with commutator (\ref{1}). In this case the $*$ product is just
the product of operators and integration over the $(X^1,X^2)$
plane is a trace,
\be
\int dx^1dx^2 f(x^1,x^2) = 2\pi \theta {\rm Tr} \hat f(X^1,X^2)
\ee
In this framework, it is convenient to introduce
complex variables $z$ and $\bar z$
\be
z = \frac{1}{\sqrt{2}}(x^1 + i x^2)\, , \;\;\;\;\;\;
\bar z= \frac{1}{\sqrt{2}}(x^1 - i x^2)
\label{2}
\ee
and annihilation and creation operators  $\hat a$ and $\hat a^\dagger$
in the form
\be
\hat a = \frac{1}{\sqrt{2\theta}}(X^1 + i X^2)\, , \;\;\;\;\;\;
\hat a^\dagger = \frac{1}{\sqrt{2\theta}}(X^1 - i X^2)
\label{3}
\ee
so that (\ref{1}) becomes
\be
[\hat a,\hat a^\dagger] = 1
\label{4}
\ee
In this way, through the action of $a^\dagger$ on the vacuum state
$|0\rangle$, eigenstates of the number operator
\be
\hat N = a^\dagger a
\label{number}
\ee
are generated.
With this conventions, derivatives in the Fock space are given by
\be
\partial_z = -\frac{1}{\sqrt \theta}
[\hat a^\dagger,~] \, , \;\;\;\;\;\; \partial_{\bar z} =
\frac{1}{\sqrt \theta} [\hat a,~] \label{5} \ee
After introducing
\be
\hat A_z = \frac{1}{\sqrt {2}} (\hat A_1 - i \hat A_2) \, ,
\;\;\;\; \hat A_{\bar z} = \frac{1}{\sqrt {2}} (\hat A_1 + i \hat
A_2) \ee
the field strength and covariant derivatives take  the form
\begin{eqnarray}
\hat F_{z \bar z}&=& \partial_z \hat A_{\bar z} - \partial _{\bar
z} \hat A_z -i[\hat A_z,\hat A_{\bar z}] \nonumber \\
&=& -
\frac{1}{\sqrt \theta} \left ([\hat a^\dagger,\hat A_z] +
[\hat a,A_{\bar z}] + i\sqrt \theta[\hat A_z,\hat A_{\bar z}]
\right) \equiv i \hat B \\
D_{\bar z}\hat \phi &=& \partial_{\bar z} \hat \phi - i A_{\bar z}
\hat \phi = \frac{1}{\sqrt \theta} [\hat a,\hat \phi] - i \hat
A_{\bar z} \hat\phi \nonumber\\ D_{z}\hp &=& \partial_z \hp+ i
\hat A_{z} \hp = -\frac{1}{\sqrt \theta} [\hat a^\dagger,\hp] + i
\hat A_z \hp
\end{eqnarray}
with $\hat B$ the magnetic field.

We will be interested in the noncommutative extension of the
nonrelativistic and relativistic  Chern-Simons-matter systems
introduced, in ordinary space, in refs.\cite{JP}-\cite{JW}. The
gauge field dynamics for these models is governed by the
Chern-Simons Lagrangian $L_{CS}[A]$ defined as
\be
L_{CS}[A] = \kappa\; \varepsilon_{\mu\nu\alpha}\left(A_\mu *
\partial_\nu  A_\alpha
- \frac{2i}{3}
A_\mu*A_\nu*A_\alpha)\right)
\ee
The Lagrangian for the noncommutative extension of the
non-relativistic case will be taken as
\be
L =  L_{CS}[A]+ i\bar \phi * D_0\phi +\frac{1}{2}\overline{D_i \phi}*
D_i\phi - \frac{1}{4}\lambda  \phi* \bar \phi *\phi * \bar \phi \label{6}
\ee
while for the relativistic case,
\be
L =  L_{CS}[A]+   \overline{D_\mu \phi}*
D^\mu\phi - V[\phi * \bar \phi] \label{6r}
\ee
with  $V$ the sixth order potential
\be
V[\phi * \bar \phi] = \frac{1}{\kappa^2} \phi*\bar \phi *
\left(\phi * \bar \phi - v^2 \right)^2
\ee
taken at the selfdual point, where Bogomol'nyi equations can be
found.

\section{BPS equations for the non-relativistic case}

The  Hamiltonian associated with Lagrangian (\ref{6})  is simply
given by
\be
H = \int d^2x \left( \frac{1}{2} \overline{D_i \phi}* D_i\phi  +
\frac{1}{4} \lambda  \phi* \bar \phi *\phi * \bar \phi \right)
\label{H}
\ee
It can be written in the form
\be
H = \int d^2x \left(- \frac{1}{2}\bar \phi  (D_1 + i \alpha D_2)
(D_1 - i \alpha D_2) \phi + \phi * \bar \phi *\left(-\frac{\alpha}{2}
F_{12} + \frac{1}{4} \lambda \phi * \bar \phi
\right)
\right)
\label{H2}
\ee
where $\alpha = \pm 1$. We shall call $\alpha = -1$ the {\it
selfdual} case and $\alpha = +1$ the {\it anti selfdual} one.

Using the Gauss law deriving from Lagrangian (\ref{6}),
\be
\kappa \varepsilon_{ij}F_{ij} + \phi * \bar \phi = 0
\label{gauss}
\ee
the Hamiltonian takes the form
\be
H = \int d^2x \left(- \frac{1}{2}\bar \phi  (D_1 + i \alpha D_2)
(D_1 - i \alpha D_2) \phi - \frac{1}{2}(\alpha +\lambda \kappa)
\phi * \bar \phi * F_{12}
\right)
\label{H3}
\ee
Then, if the following relation among the two free parameters in
the theory holds
\be
\lambda \kappa = -\alpha
\label{rel}
\ee
the lower bound for the Hamiltonian is attained when the following
Bogomol'nyi equations are satisfied
\begin{eqnarray}
&&\left(D_1 - i\alpha D_2\right) \phi = 0 \nonumber\\
&&B = -\frac{1}{2\kappa} \phi * \bar \phi
\label{bog}
\end{eqnarray}

Let us first consider the selfdual ($\alpha = -1$) case. In
operator language,  equations (\ref{bog}) can then be written as
\begin{eqnarray}
&&D_{\bar z}\hat \phi = 0 \nonumber\\
&&B = -\frac{1}{2\kappa} \hat \phi   \hp
\label{bogo}
\end{eqnarray}
In order to search for vortex solutions to these equations, we
propose the  ansatz
\begin{eqnarray}
\hat \phi &=& \sqrt{\frac{2|\kappa|}{\theta}}\;
\sum_{n=0}^\infty f_n |n\rangle
\langle n +
M-1|\label{union}\\
\hat A_z &=&  \frac{i}{\sqrt{\theta}}
\sum_{n=0}^\infty d_n |n+1\rangle \langle n |
\label{uni}
\end{eqnarray}
where $f_n$ and $d_n$ are arbitrary real coefficients and
$\{|n\rangle\}$ is the basis provided by the number operator $\hat
N$. The ansatz (\ref{union}) leads, in the $\theta \to 0$ limit,
to $\phi \sim \rho(r) z^{M-1}$ which corresponds, in ordinary
space, to the usual cylindrically symmetric ansatz with a Higgs
field phase $(M-1)\varphi$ \cite{JP}.

Inserting ansatz (\ref{union})-(\ref{uni}) into eq.(\ref{bogo})
one obtains the following recurrence relations
\begin{eqnarray}
&&2\sqrt p \, d_{p-1} - d^2_{p-1} - 2 \sqrt{p+1}\, d_p + d_p^2 =
-\frac{|\kappa|}{\kappa} f_{p}^2\nonumber\\
&& d_p = \sqrt{p+1} - \sqrt{p+M}\frac{f_p}{f_{p+1}}
\label{recu1}
\end{eqnarray}
which can be combined into the following recurrence relation for the
$f_n's$ coefficients
\begin{eqnarray}
f_1^2 &=& \frac{M f_0^2}{1 - \frac{|\kappa|}{\kappa} f_0^2}
\nonumber\\
f_{p+1}^2 &=&  \frac{(p+M)f_p^2}
{1-\frac{|\kappa|}{\kappa} f_p^2 + (p+M-1)
f_{p-1}^2/f_p^2}\, , \; \; \; \; p \geq 1
\end{eqnarray}

If, instead, we choose the anti selfdual case ($\alpha = 1$), the
equations to solve read
\begin{eqnarray}
&&D_{ z}\hat \phi = 0 \nonumber\\
&&B= -\frac{1}{2\kappa} \hat \phi   \hp
\label{bogo2}
\end{eqnarray}
In this case, the appropriate ansatz is
\begin{eqnarray}
\hat  \phi &=& \sqrt{\frac{2|\kappa|}{\theta}}
\sum_{n=0}^\infty f_n |n+M-1\rangle
\langle n |\nonumber\\
\hat  A_z &=&  \frac{i}{\sqrt{\theta}}
\sum_{n=0}^\infty d_n |n+1\rangle \langle n |
\label{dosi} \end{eqnarray}
and the recurrence relations become
\begin{eqnarray}
&&
2\sqrt p \, d_{p-1} - d^2_{p-1} - 2 \sqrt{p+1}\, d_p + d_p^2 =
-\frac{|\kappa|}{\kappa} f_{p-1}^2 \nonumber \\
&&d_{p+1} = \sqrt{p+M} - \sqrt{p+1}\frac{f_{p+1}}{f_{p}}
\label{recu2}
\end{eqnarray}
combining to
\begin{eqnarray}
f_1^2 &=&
\left(M - \frac{|\kappa|}{\kappa} f_0^2\right)f_0^2
\nonumber\\
f_{p+1}^2 &=& \frac{1}{p+1} f_{p}^2
\left( -\frac{|\kappa|}{\kappa}f_{p}^2 +1 +
p
\frac{f_{p}^2}{f_{p-1}^2}
\right)\, , \; \; \; \; p \geq 1
\end{eqnarray}
The flux of the  solutions is given by
\be
\frac{\Phi}{2\pi} = \theta {\rm Tr} B = - \frac{|\kappa|}{\kappa}
\sum_p f^2_p
\ee
Finite flux configurations correspond to solutions such that
\begin{equation}
\lim_{p\to\infty}f_p=0
\end{equation}

The analysis of the asymptotic behavior of the recurrence
relation (\ref{recu1}) shows
that, for large $p$,
\begin{equation}
f^2_p \rightarrow \frac{1}{p^{\beta}} \;\;\;\;\;  \label{cero}
\end{equation}
with $\beta$ a real positive parameter to be determined. The flux
can also be obtained directly from the expression of $B$ as
\begin{equation}
\frac{\Phi}{2\pi}=\lim_{p\to\infty} \left(d_p^2-2d_p \sqrt{p+1}\right)
\end{equation}
Using eqs. (\ref{recu1})-(\ref{recu2}) one gets,
\begin{eqnarray}
\frac{\Phi}{2\pi}=M-1+\beta   & & \kappa<0 \label{f1}\\
\frac{\Phi}{2\pi}=-M+1-\beta   & & \kappa>0
\label{f2}
\end{eqnarray}

The numerical study of the recurrence relations reveals that
eq.(\ref{recu1}) has solutions only for   $\kappa<0$ while
eq.(\ref{recu2}) has solutions only for $\kappa>0$. Thus, as in
the commutative case,  solutions exist only for $\lambda<0$, that
is, only for an ``attractive'' interaction. Given an initial
arbitrary value $f^2_0$ for the recurrence relation, all
coefficients can be determined in such a way that they satisfy
(\ref{cero}). Since $\theta$ does not enter explicitly in the
recurrence relations, one can construct a whole family of
solutions parametrized by $\theta$ (which appears as a factor, see
eqs.(\ref{uni}) or (\ref{dosi})).

We have explored numerically the whole $f^2_0 \geq 0$ range
finding, for the selfdual case, that there exists, for every value
of $f_0$, a consistent solution. In contrast, in the anti-selfdual
case, the solution ceases to exist for $f_0^2 > M$. This is
reminiscent of what happens for non-commutative Nielsen-Olesen
anti-selfdual vortices \cite{BLP}.

We have find that $\beta$ is a monotonically increasing function
of $f_0^2$ such that
\be
\lim_{f_0 \to 0}\beta = M+1
\ee
Now, according to eqs.(\ref{f1})-(\ref{f2}) the magnetic flux of
our exact solution is in general not quantized in non-commutative
space.

We show in figure 1 and 2 the magnetic field $B$  as a function of
$r$, for the selfdual and anti-selfdual solutions respectively,
computed from eqs.(\ref{bogo}),(\ref{bogo2}).  We plot  different
values of $f_0^2$ for a given $\theta$. One sees that the  $B$ is
reminiscent of the magnetic field corresponding to the ordinary
(commutative) case, except that in the latter case $B(0)= 0$ while
in the present noncommutative case $B(0) = B_0$, with $B_0$
positive (negative) for the selfdual (anti-selfdual) case.

Let us relate at this point  this result with that corresponding
to ordinary space. As originally shown in \cite{JP} Bogomol'nyi
equations for the non-relativistic system can be exactly solved
since the problem can be reduced to finding solutions to the
Liouville equation. Then, the most general axially symmetric
regular solution gives, for the selfdual case,
\be
\phi^{comm} = \frac{2\sqrt{2|\kappa|}}{r} M
\left( \left(\frac{r_0}{r}\right)^M + \left(\frac{r}{r_0}\right)^M
\right) ^{-1}\!\!\exp\left(i(M-1)\,\varphi\right)
\; , \;\;\;\; M=1,2, \ldots
\label{solv}
\ee
where $r_0$ is an integration constant. The other free parameter is
$M$ which is quantized on regularity grounds. Accordingly, the
magnetic flux associated to this solution is quantized,
\be
\Phi^{comm} = 2\pi   (2M)
\label{viejo}
\ee
Note that the flux of our selfdual noncommutative solutions
coincides, in the ${f_0 \to 0}$ limit with that of the ordinary
case. To study the connection between our solution and that in
ordinary space in more detail, let us consider the $\theta \to 0$
limit of the former in configuration space. It is enough to
consider the small $r$ region where the commutative solution
(\ref{solv}) can be written in the form
\be
\phi^{comm} = \frac{2\sqrt{2|\kappa|} }{r_0^M} M z^{M-1} + O(r^{3M-1})
\label{eqo}
\ee
Since, for small $\theta$, $r^2 \approx \theta \hat N$, the
leading contribution in the noncommutative case corresponds to the
$n=0$ term in solution (\ref{union}),
\begin{eqnarray}
\hat \phi &\approx&
\sqrt{\frac{2|\kappa|}{\theta}} f_0|0\rangle \langle M-1| =
\sqrt{\frac{2|\kappa|}{\theta}} f_0|0\rangle \langle 0|
\frac{a^{M-1}}{\sqrt{(M-1)}!}\nonumber\\
& \approx&
\sqrt{\frac{2|\kappa|}{\theta}} f_0|0\rangle \langle 0|
\frac{z^{M-1}}{\sqrt{(2\theta)^{M-1}(M-1)}!}
\label{eqi}
 \end{eqnarray}
A relation between
$r_0$ and $f_0$ can be found comparing eqs.(\ref{eqo}) and (\ref{eqi})
\be
f_0^2 = 2^{M+1} {M!M} \left(\frac{\theta}{r_0^2}\right)^M
\label{sati}
\ee
If $f_0^2$ does not vanish as $\theta^M$ in the $\theta \to 0$
limit, the noncommutative solutions goes in this limit to a
singular solution in ordinary space. Only when the behavior
(\ref{sati}) is satisfied, the $\theta \to 0$ limit converges to
the Jackiw-Pi solution \cite{JP}. We have numerically checked this
finding that, already for $\theta/r_0^2 \approx 0.01$, the
noncommutative and the Jackiw-Pi solutions are indistinguishable.

\section{BPS equations for the relativistic model}

The associated Hamiltonian for the model (\ref{6r}) for static
field configurations is
\be
H = \int d^2x \left(    \overline{D_i \phi}* D_i\phi  +
A_0*A_0* \phi *\bar \phi + V[\phi * \bar \phi] \right)
\label{ha}
\ee
The Gauss law deriving from (\ref{6r}) takes, for static
configurations, the form
\be
2\kappa B = - \left(
\phi * \bar \phi* A_0 + A_0*\phi*\bar \phi
\right)
\label{relgauss}
\ee
Assuming that $A_0$ (Moyal) commutes with $\phi * \bar \phi$ (as
it will be the case for our ansatz, see below) we have
\be
A_0 = -\kappa  ({\phi * \bar \phi})^{-1} * B
\label{asero}
\ee
Inserting (\ref{asero}) in (\ref{ha}) and integrating by parts one
gets
\begin{eqnarray}
H &=&\int d^2x \left(\vphantom{\frac{1}{2\kappa^2}^2}
\overline{(D_1 + i \alpha D_2)
\phi} *
(D_1 - i \alpha D_2) \phi   + \right.\nonumber\\
&&\left. \frac {\kappa^2}{ |\phi|^2} *
\left( B + \alpha \frac{1}{2\kappa^2}|\phi|^2*(|\phi|^2 - v^2)
\right)^2
\right) - \alpha v^2 \Phi
\label{hami}
\end{eqnarray}
where $ \alpha = \pm 1$, $|\phi|^2 = \phi* \bar \phi$ and $\Phi =
\int d^2x B$ is, as before, the magnetic flux. Thus, in the {\it
selfdual case} ($\alpha=-1$) the energy is bounded by $v^2 \Phi$,
and the bound is saturated when the selfdual equations are
fulfilled
\begin{eqnarray}
&& D_{\bar z} \phi = 0 \; , \nonumber \\
&& B +\frac{1}{2\kappa^2}|\phi|^2*(|\phi|^2 - v^2)=0
\label{rel-sd}
\end{eqnarray}
Analogously, in the {\it anti-selfdual} case ($\alpha = 1$) the
energy bound is $- v^2 \Phi$ and is reached when the anti-selfdual
equations are satisfied
\begin{eqnarray}
&& D_{z} \phi = 0 \; , \nonumber \\
&& B -\frac{1}{2\kappa^2}|\phi|^2*(|\phi|^2 - v^2)=0
\label{rel-antisd}
\end{eqnarray}

Let us analyze the selfdual case first. As in the non-relativistic
case we shall work in the operator framework and propose an ansatz
of the form
\begin{eqnarray}
\hat \phi &=& v\sum_{n=0}^\infty f_n |n\rangle
\langle n +
M|\label{union2}\\
\hat A_z &=&  \frac{i}{\sqrt{\theta}}\sum_{n=0}^\infty d_n
|n+1\rangle \langle n |
\label{ansatz-rel}
\end{eqnarray}
where again  $f_n$ and $d_n$ are arbitrary real coefficients. With
this ansatz both the magnetic field and $|\phi|^2$ are diagonal
\begin{eqnarray}
B &=& \frac{1}{\theta} \left( \sum_{n=1}
\left( g_{n-1} - g_m \right) |n\rangle
\langle n | - g_0 |0\rangle\langle0| \right) \label{magne} \\
|{\hat \phi}|^2 &=& v^2 \sum_{n=0} f_n^2 |n\rangle \langle n |
\label{b}
\end{eqnarray}
where $g_n = 2 d_n \sqrt{n+1} - d_n^2$. Consequently, from the
Gauss law (\ref{relgauss}), we see that $A_0$ commutes with $|\hat
\phi|^2$ and can be solved as in equation (\ref{asero}).

The selfdual system (\ref{rel-sd}) is then equivalent to the
following system of recurrence relations
\begin{eqnarray}
d_p &=& \sqrt{p+1} - \sqrt{p+M}\frac{f_p}{f_{p+1}}\, ,
\; \; \; \; p\geq 0  \nonumber\\
f_1^2 &=& \frac{(M+1) f_0^2}{1 + a\; f_0^2 (1 - f_0^2)}
\nonumber\\
f_{p+1}^2 &=&  \frac{(p+M +1)f_p^2}{1 + a\; f_p^2 (1- f_0^2) + (p+M)
f_{p-1}^2/f_p^2}\, , \; \; \; \; p\geq 1
\label{sys1}
\end{eqnarray}
where $a = v^4 \theta/(2 \kappa^2)$.

For the anti-selfdual case, the ansatz we propose is
\begin{eqnarray}
\hat \phi &=& v\sum_{n=0}^\infty f_n |n + M\rangle
\langle n |\label{union3}\\
\hat A_z &=&  \frac{i}{\sqrt{\theta}}\sum_{n=0}^\infty d_n
|n+1\rangle \langle n |
\label{ansatz-rel2}
\end{eqnarray}
The magnetic field has the same form as in (\ref{b}) while one has for
the scalar field
\begin{eqnarray}
|{\hat \phi}|^2 &=& v^2 \sum_{n=0} f_n^2 |n +M\rangle \langle n +M |
\label{b2}
\end{eqnarray}
The anti-selfdual recurrence relation take the form
\begin{eqnarray}
d_{p+1} &=& \sqrt{p+M+1} - \sqrt{p+1}\frac{f_{p+1}}{f_{p}}
\nonumber\\
f_1^2 &=&
\left(M+1 - a f_0^2(1-f_0^2)\right)f_0^2
\nonumber\\
f_{p+1}^2 &=& \frac{1}{p+1} f_{p}^2
\left(1 -a f_{p}^2(1-f_{p}^2) +
p
\frac{f_{p}^2}{f_{p-1}^2}
\right)\, , \; \; \; \; p \geq 1
\label{sys2}
\end{eqnarray}

We have studied systems (\ref{sys1}) and (\ref{sys2}) numerically.
Given a value for $f_0$  one can then determine all $f_n$'s from
(\ref{sys1}) or (\ref{sys2}). The correct value  for $f_0$  should
make $f_n^2 \to 1$ asymptotically  so that boundary conditions are
satisfied (we are looking for symmetry breaking solutions). The
values of these coefficients will depend on the choice of the
dimensionless parameter $a =\theta v^4/(2\kappa^2)$. We have
explored the whole range of $a$ and found a consistent solution
for any positive integer $M$ both in the selfdual and in the
anti-selfdual case. In contrast with the non-relativistic model,
there is only one value of $f_0$ leading to the appropriate
boundary condition, for each value of $a$. For example, for the
self-dual case we have
\begin{eqnarray}
a = 0.5, & & f_0^2 = 0.2168142\ldots \nonumber\\
a = 1.0, & & f_0^2 = 0.4037747\ldots \nonumber\\
a = 2.0, & & f_0^2 = 0.6228436\ldots
\end{eqnarray}

Once the $f_n's$ and $d_n's$ are  determined in this way, the
magnetic field can be computed using eq.(\ref{magne}) or
Bogomol'nyi equation. As an example, for the selfdual case, one
has
\be
\hat B = \frac{v^2}{2\kappa^2}
 \sum_{n=0}^{\infty} f_n^2\left(1 - f_n^2\right) \vert n
\rangle \langle n \vert \label{ayux} \ee
or, using the explicit formula for $\vert n \rangle \langle  n
\vert $ in configuration space \cite{GMS}
\be
B(r) = \frac{v^2}{\kappa^2} \sum_{n=0}^{\infty}(-1)^n
f_n^2\left(1 - f_n^2\right)
\exp(-\frac{r^2}{\theta}) L_n(2\frac{r^2}{\theta}) \label{Bx}
\ee
where $L_n$ are the Laguerre polynomials.

We show in figure 3 (selfdual case) and 4 (anti-selfdual case) the
resulting magnetic field $B$ as a function of $r$ for different
values of $\theta$. For $\theta = 0$ we recover in both cases the
CS vortex solutions found in \cite{HKP}-\cite{JW}. One should note
that, as in the non-relativistic case, the magnetic field profile
corresponding to the anti-selfdual case is not the trivial reverse
of the selfdual one. This is related, as before, to the presence
of the parity breaking parameter $\theta$.  As $\theta$  grows,
the magnetic field differs more and more from the annulus-shaped
ordinary CS vortex with a value at the origin which grows till
$B(0)$ becomes a maximum. It is important to stress that we have
found vortex solutions in the whole range of $\theta$ {\it both in
the selfdual and anti-selfdual cases} in contrast with what
happens for Nielsen-Olesen vortices where anti-selfdual solutions
do not exist for  $\theta$ larger than a critical value
\cite{BLP}.

The magnetic flux of the  solutions  can be computed  using
\be
\Phi = 2 \pi \theta {\rm Tr} \hat B
\label{flujo}
\ee
One finds,
\be
\Phi = 2\pi M \, , \;\;\;\; M=1,2, \ldots
\label{quantica}
\ee
showing that in the relativistic case, the magnetic flux is
quantized for all $\theta$. This, and the expression (\ref{hami})
for the Hamiltonian allows to write the energy of the selfdual
($\alpha = -1$) and anti-selfdual ($\alpha = 1$) solitons in the
form
\be
H =  (2\pi v^2)  M \label{cota}
\ee
which coincides with the expression for CS solitons in ordinary
space first found in \cite{HKP}-\cite{JW}

\section{Conclusions}

In summary, we have constructed exact soliton solutions to
non-commutative Chern-Simons theory coupled to a charged scalar in
3-dimensional space-time. We have shown the existence of first
order Bogomol'nyi equations and we have found, in the
non-relativistic case,  that an attractive $\phi^4$ interaction
guarantees, as in ordinary space, the existence of regular
vortex-like solutions. There is however an important difference
that manifests at finite $\theta$: while non-topological ordinary
solitons have, on regularity grounds, an associated quantized
magnetic flux, noncommutative solitons can have arbitrary flux.
Only in the $\theta \to 0$ limit, in which these solutions
approach smoothly the ordinary ones, the flux becomes quantized.
Remarkably, one can find a relation between the arbitrary
integration constant arising in the solution of the Liouville
equation satisfied by the Higgs field in ordinary space and the
noncommutative parameter $\theta$. It should be stressed that
because of the presence of $\theta$, anti-selfdual solutions can
not be trivially obtained from selfdual ones by making $B \to -B$.
In fact, we have shown that although solutions exist in both
cases, there is a range of parameters where anti-selfdual solitons
cease to exist. Concerning the relativistic case, as in ordinary
space, a $\phi^6$ potential guarantees the existence Bogomol'nyi
equations and  vortex-like topological solutions which also
approach smoothly ordinary ones when $\theta \to 0$. Again, the
presence of $\theta$ make selfdual and anti-selfdual solutions not
trivially connected. Expressions for the magnetic flux and the
energy of the CS solitons coincide with those in ordinary space
for arbitrary $\theta$.

As stressed in the introduction, one of the interests in CS
solitons concerns their possible use in understanding relevant
phenomena  in planar physics. The connection between
non-commutative field theories and systems in strong magnetic
fields \cite{SJ}-\cite{BS} make them attractive for a field
theoretical approach to the Quantum Hall and Bohm-Aharonov effects
\cite{Pa}-\cite{Shu}. We hope to report on these issues in a
separate publication.

\vspace{1 cm}

\noindent\underline{Acknowledgements}: This work is partially
supported by CICBA, CONICET (PIP 4330/96), ANPCYT (PICT 97/2285).
G.S.L.  and E.F.M. are partially supported by Fundaci\'on
Antorchas.


\newpage


~

\begin{figure}
\vspace{-9. cm}
\hspace{-0.34cm}{\centerline{ \psfig{figure=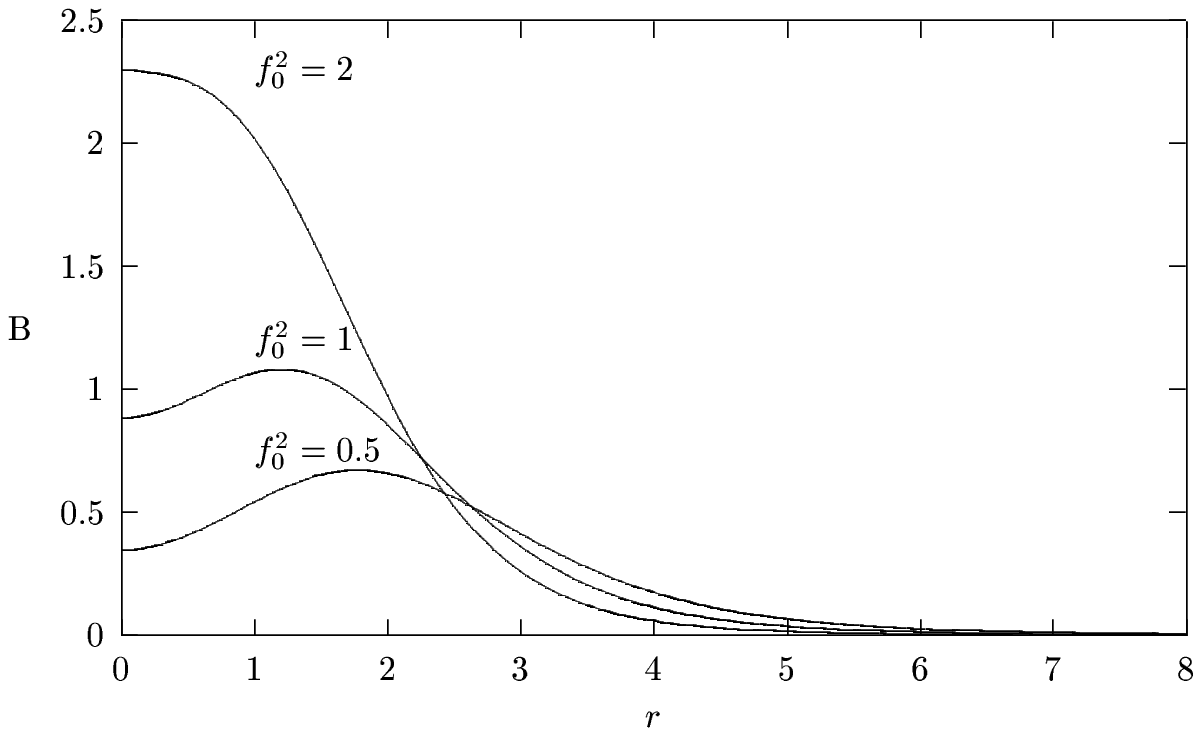,angle=0}}}
\vspace{-12.cm}
\caption{ The magnetic field $B(r)$ for the
selfdual non-relativistic solution for different values of $f_0$.
$\theta$ and $\kappa$ have been taken equal to 1.}
\end{figure}

~

\begin{figure}
\vspace{-16. cm}
\hspace{-0.34cm}{\centerline{ \psfig{figure=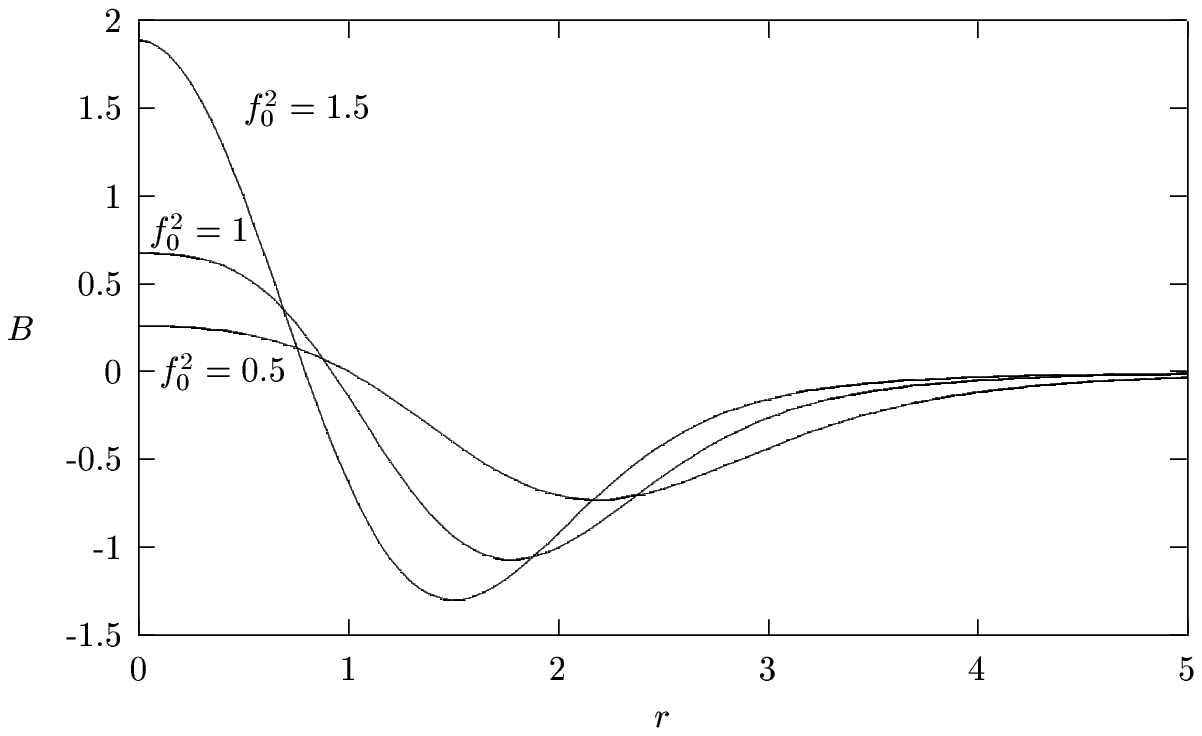,angle=0}}}
\vspace{-12.cm}
\caption{ The magnetic field $B(r)$ for the
anti-selfdual non-relativistic solution for different values of $f_0$.
$\theta$ and $\kappa$ have been taken equal to 1.}
\end{figure}

\newpage

~

\begin{figure}
\vspace{-9. cm}
\hspace{-0.34cm}{\centerline{ \psfig{figure=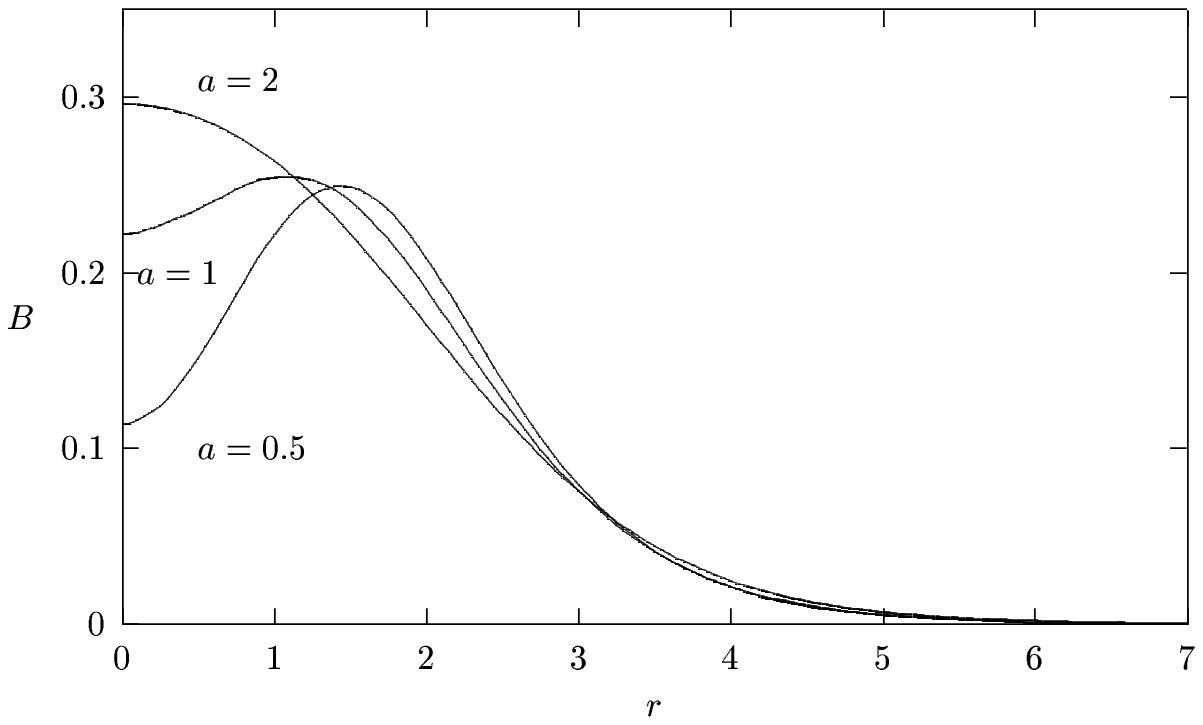,angle=0}}}
\vspace{-12.cm}
\caption{ The magnetic field $B(r)$ for the
selfdual relativistic solution for different values of
$a = v^2\theta/(2\kappa^2)$.}
\end{figure}

\begin{figure}
\vspace{-16. cm}
\hspace{-0.34cm}{\centerline{ \psfig{figure=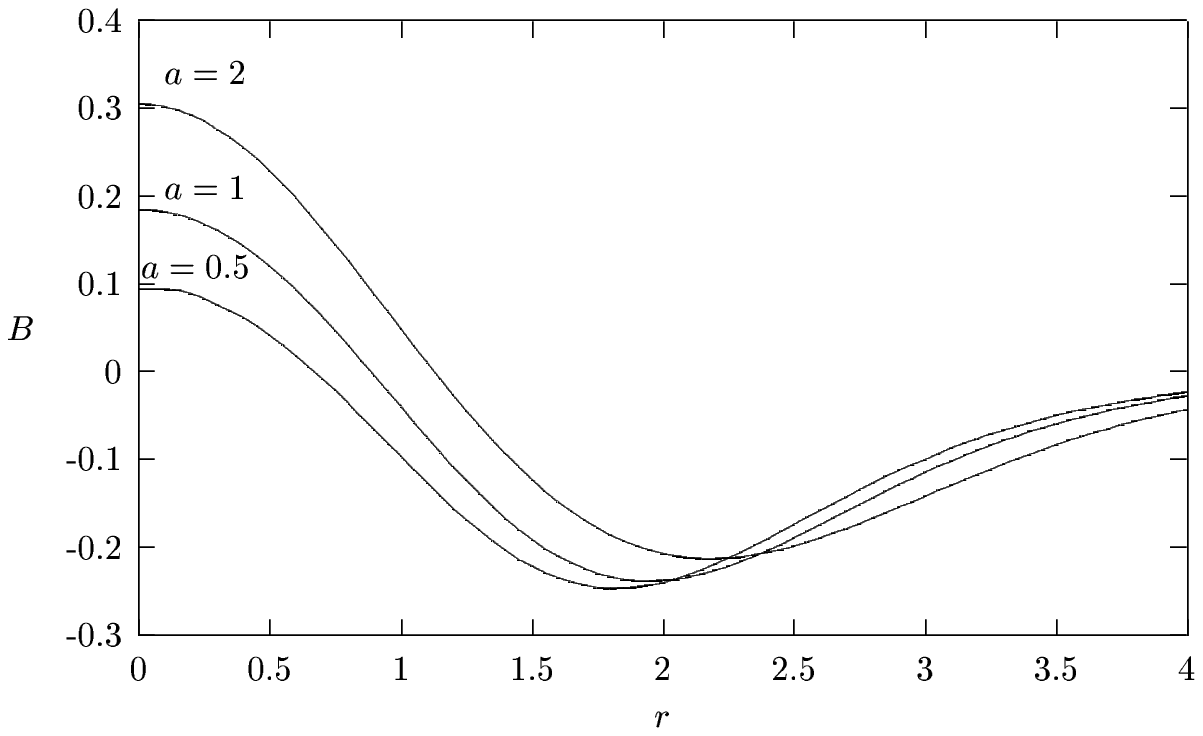,angle=0}}}
\vspace{-12.cm}
\caption{ The magnetic field $B(r)$ for the
anti-selfdual relativistic solution for different values of
$a = v^2\theta/(2\kappa^2)$.}
\end{figure}



\begin{thebibliography}{99}
\bibitem{CDS} A.Connes, M.R.~Douglas and A.S.~Schwarz,
JHEP {\bf 02} (1998) 003.
\bibitem{DH} M.R.~Douglas and C.~Hull, JHEP {\bf 02} (1998) 008.
\bibitem{SW} N.~Seiberg and E.~Witten, JHEP {\bf 09} (1999) 032.
\bibitem{HA} A.~Hashimoto, JHEP {\bf 9911}
(1999) 005.
\bibitem{Mr} S.~Moriyama, Phys.Lett. {\bf B485} (2000)
278.
\bibitem{GMS} R.~Gopakumar,
 S.~Minwalla and A.~Strominger, JHEP {\bf 0005}
(2000) 020.
\bibitem{GN} D.J.~Gross and N.~Nekrasov, JHEP {\bf 0007} (2000) 034;
 JHEP {\bf 0010} (2000) 021; hep-th/0010090.
\bibitem{JMW} D.P.~Jaktar, G.~Mandal and S.R.~Wadia, JHEP {\bf 0009}
 (2000) 018.
 \bibitem{Gorsky} A.S. Gorsky, Y.M. Makeenko, K.G. Selivanov
 Phys.Lett. {\bf B492} (2000) 344.
\bibitem{AGMS} M.~Aganagic, R.~Gopakumar, S.~Minwalla and A.~Strominger,
hep-th/0009142
\bibitem{P} A.P.~Polychronakos, Phys. Lett. {\bf B495} (2000) 407.
\bibitem{NK} N.~Nekrasov,  hep-th/0010017.
\bibitem{JKL} J.A.~Harvey, P.~Kraus and F.~Larsen,
hep-th/0010060.
\bibitem{Te} M.~Hamanaka and S~Terashima, hep-th/0010221
\bibitem{KH} K.~Hashimoto,  hep-th/0010251.
\bibitem{Bk} D.~Bak, hep-th/0008204.
\bibitem{BLP} D.~Bak, K.~Lee and J-H.~Park, hep-th/0011099.
\bibitem{LMS} G.S.~Lozano, E.F.~Moreno and F.A.~Schaposnik,
Phys. Lett. {\bf B} (in press)  hep-th/0011205.
\bibitem{Fr} See E.~Fradkin, {\it Field Theories of
Condensed Matter systems}, and references therein.
\bibitem{J} See R.Jackiw in {\it Diverse Topics in
Theoretical and Mathematical Physics}  World Sci.
Singapore, 1995, pp 465-514 and references therein.
\bibitem{Chu} C.~Chu,  Nucl. Phys. {\bf B580} (2000) 352.
\bibitem{BGPS} A.~A.~Bichl, J.~M.~Grimstrup, V.~Putz and M.~Schweda,
JHEP{\bf 0007} (2000) 046.
\bibitem{CW}  G.~Chen and Y.~Wu, hep-th/0006114.
\bibitem{M} S.~Mukhi and N.~V.~Suryanarayana, JHEP{\bf 0011} (2000) 006.
\bibitem{GS} N.~Grandi and G.~A.~Silva, hep-th/0010113.
\bibitem{Pol2} A.~P.~Polychronakos, JHEP{\bf 0011} (2000) 008.
\bibitem{JMW2} V.~John, A.~V.~Nguyen and C.~W.~Kameshwar,
Phys. Lett. {\bf B371} (1996) 252.
\bibitem{JP} R,~Jackiw and S.-Y.~Pi, Phys. Rev. Lett. {\bf 64} (1990) 2969,
{\bf (C)} {\bf 66} (1991) 2682; Phys. Rev. {\bf D42} (1990) 3500.
\bibitem{HKP} J.~Hong, Y.~Kim and P.Y.~Pac, Phys. Rev. Lett.
{\bf 64} (1990) 2230.
\bibitem{JW} R.~Jackiw and E.~Weinberg, Phys. Rev. Lett.
{\bf 64} (1990) 2234;
R.~Jackiw, K.~Lee and E.~Weinberg, Phys. Rev. {\bf D42} (1990) 3488.
\bibitem{dVS} H.~de Vega and F.A.~Schaposnik, Phys. Rev. {\bf D14} (1976)
1100.
\bibitem{Bogo} E.B.~Bogomol'nyi, Sov. Jour. Nucl. Phys. {\bf 24} (1976) 449.
\bibitem{SJ} M.~Sheikh-Jabbari, Phys. Lett. {\bf B455} (1999) 129.
\bibitem{BS} D.~Bigatti and L.~Susskind, Phys. Rev. {\bf D62} (2000) 066004.
\bibitem{Pa} V. Pasquier, Phys.Lett. {\bf B490} (2000) 258.
\bibitem{BBST} B.A.~Bernevig, J.~Brodie, L.~Susskind and N.~Toumbas,
hep-th/0010105.
\bibitem{GR} S.S.~Gubser and M.~Rangamani, hep-th/0012155.
\bibitem{L} G.~Lozano, Phys. Lett. {\bf B283} (1992) 70;
 O~.Bergmann and G.~Lozano, Ann. Phys. (N.Y.) {\bf 229} (1994)
416.
\bibitem{Shu} M.~Chaichian, A.~Demichev, P.~Pre{s}najder, M.M.~Sheikh-Jabbari
and A.~~Tureanu, hep-th/0012175.
\end{thebibliography}
\end{document}